# Stopping light by an air waveguide with anisotropic metamaterial cladding

**Tian Jiang, Junming Zhao and Yijun Feng***

*Key Laboratory of Modern Acoustics, Ministry of Education, and Department of Electronic Science and Engineering, Nanjing University, Nanjing, 210093, China*
*\*Corresponding author: yjfeng@nju.edu.cn*

**Abstract:** We present a detailed study of oscillating modes in a slab waveguide with air core and anisotropic metamaterial cladding. It is shown that, under specific dielectric configurations, slow and even stopped electromagnetic wave can be supported by such an air waveguide. We propose a linearly tapped waveguide structure that could lead the propagating light to a complete standstill. Both the theoretical analysis and the proposed waveguide have been validated by full-wave simulation based on finite-difference time-domain method.

## 1. Introduction

Devices that can buffer or delay the light pulse are key components for both optical communication networks and quantum information processing systems. Large pulse delay (or 'slow light') is achievable with many methods/structures including electromagnetically induced transparency [1], quantum-dot semiconductor optical amplifiers [2], photonic crystal waveguide [3], direct coupled resonators [4], coherent population oscillations [5], stimulated Brillouin scattering [6], stimulated Raman scattering [7] and surface plasmon polaritons [8-10]. Recently, several groups proposed the slow light structures based on metamaterials [11-15]. K. L. Tsakmakidis *et al.* bring together the realms of metamaterials and slow light by studying a three-layer slab heterostructure composed of negative-index ('left-handed material') core and positive-index ('right-handed material') cladding [13]. The physics of slowing down the group velocity is been explained by negative Goos-Hänchen lateral displacements on the media interfaces. They discover that for different excitation frequencies the guided oscillatory fields stop at correspondingly different waveguide thicknesses in a linearly tapered waveguide, which is the so-called 'trapped rainbow' effect. However, fabricating an isotropic left-handed material (LHM) at optical frequencies is a significant challenge, while several anisotropic metamaterials have been successfully demonstrated by experiments [16,17]. Moreover, the metamaterials used in the waveguide are always lossy and the attenuation of the slow light will be obvious due to the interaction between the guided light and the lossy core materials. In our previous work [18], we have proposed a planar air waveguide with anisotropic metamaterial (AMM) cladding to slow down and even trap the light. Similar structure based on negative-refractive-index photonic crystal was also proposed [19].

In this letter, we report our intensive investigation on a symmetric slab waveguide where an air core is surrounded by an AMM cladding. It is shown that the air waveguide could support different oscillating modes. By deriving closed-form expressions for the cutoff conditions of the oscillating modes, we prove that, with careful choice of the material parameters, there exist guiding modes for which extremely low or even zero group velocity can be attained. We also demonstrate by finite-difference time-domain (FDTD) method that a linearly tapped waveguide can be used to efficiently and coherently bring light pulses to a complete standstill.

## 2. Air waveguide design and mode analysis

The geometry of the waveguide is shown in Fig. 1, where an air slab of thickness $d$ is surrounded by the AMM with partially negative permittivity and/or permeability components. To simplify the analysis, we assume the permittivity and permeability tensors that are simultaneously diagonal as $\bar{\bar{\varepsilon}} = diag\left[\varepsilon_x, \varepsilon_y, \varepsilon_z\right]$, $\bar{\bar{\mu}} = diag\left[\mu_x, \mu_y, \mu_z\right]$, with their principal axis paralleling the coordinate axis used in Fig. 1. We consider primarily *p*-polarized

(transverse magnetic, TM) waves, in which the magnetic field is along the $y$ axis. Similar analysis can be carried out for the transverse electric modes. From the Maxwell equations and the boundary conditions, we could obtain the dispersion equation that defines a set of allowed oscillating modes:

$$\tan(\kappa d) = \frac{2\eta}{1-\eta^2}, \qquad (1)$$

where the following ratios are introduced: $\eta = \varepsilon_0 p/\varepsilon_z \kappa$, $ip$ and $\kappa$ standing for the transverse component of the wave vector in the AMM and the air core. Complex guided modes, i.e. modes with a complex propagation constant, have been observed in some completely lossless wave guiding structures [20,21]. To giving a complete mode spectrum, we define the complex propagation constant along z direction $k_z = \beta + i\alpha$, and the effective index $n_{eff} = (\beta + i\alpha)/k_0$, here $k_0$ representing the free-space wave number. Therefore $p$ and $\kappa$ can be obtained from the dispersion relations of the TM waves in air and AMM, i.e., $p = k_0 \left( n_{eff}^2 \varepsilon_z / \varepsilon_x - \varepsilon_z \mu_y / \varepsilon_0 \mu_0 \right)^{1/2}$, $\kappa = k_0 \left( 1 - n_{eff}^2 \right)^{1/2}$, respectively. If $\alpha \neq 0$, $p$ and $\kappa$ are assumed to be complex numbers. For each mode and different material parameters, the dispersion diagrams, $n_{eff}$ versus the reduced slab thickness $k_0 d$, may be computed by numerical method.

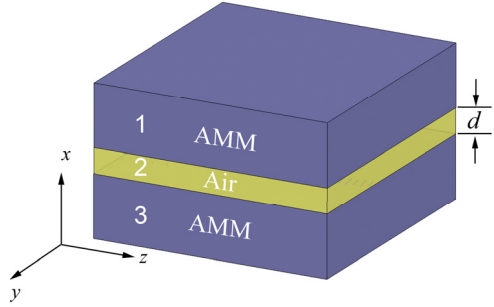

Fig. 1 A planar air waveguide with anisotropic metamaterial cladding.

Table 1. Six parameter sets of AMM supporting TM oscillating modes.

|  |  | $\varepsilon_x > 0$ | $\varepsilon_x < 0$ |
|---|---|---|---|
| $\varepsilon_z > 0$ | $\mu_y > 0$ | Set A |  |
|  | $\mu_y < 0$ | Set B | Set C |
| $\varepsilon_z < 0$ | $\mu_y > 0$ | Set D | Set E |
|  | $\mu_y < 0$ |  | Set F |

For wave propagation in the air waveguide, the $z$ component of the Poynting vector is:

$$S_z = \frac{1}{2}\text{Re}(E_x H_y^*) = \frac{\beta}{2\omega \varepsilon_{x,0}} |H_y|^2. \qquad (2)$$

Integrating $S_z$ over the waveguide's cross section, we obtain the power flow $P_i$ in different layers and the total power flow $P_{tot} = \sum_{i=1}^{3} P_i$. It is inferred from Eq. (2) that the net power flow can become negative in the AMM cladding when $\varepsilon_x < 0$, but remains positive in the air core. If $P_{tot} < 0$, the mode is a backward mode since the total energy flow is opposite to the phase velocity. Otherwise, the mode is a forward mode. Under specific conditions the energy flow in the AMM cancels out that in the air core, i.e., $P_{tot} = 0$. One can prove that the group velocity reduce to zero simultaneously. Although the above equation of the total power flow is more complicated when complex guided mode appears, the result turns out to be simply zero [21].

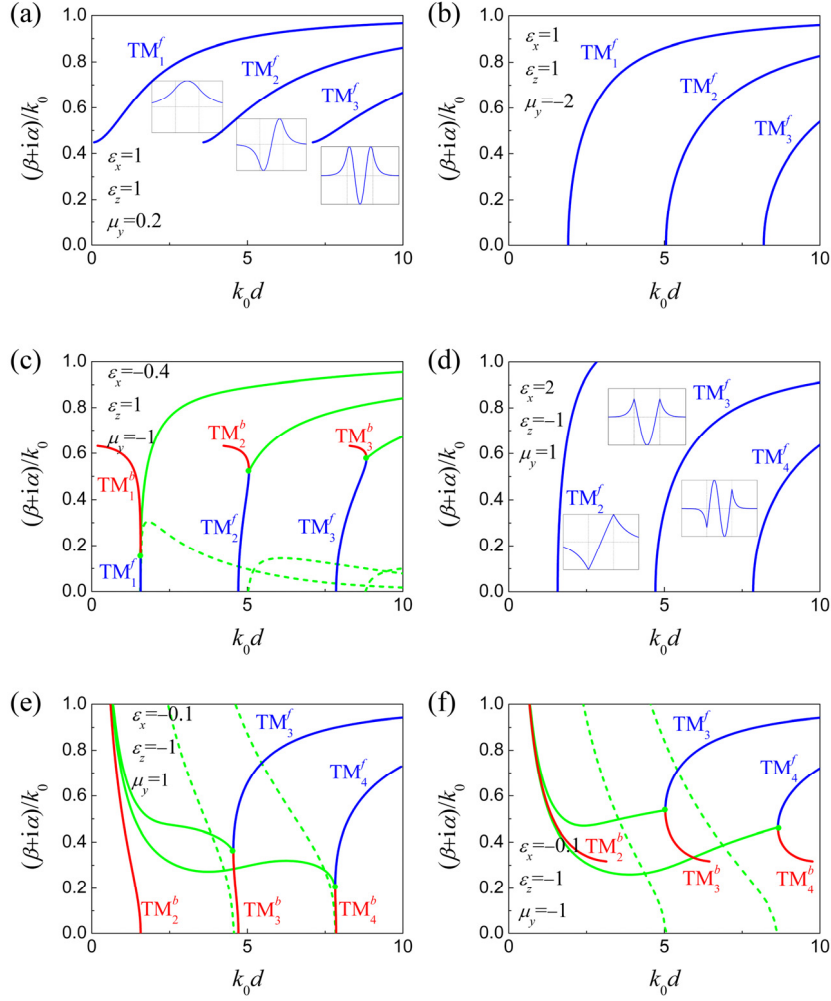

Fig. 2 The complete dispersion diagrams of the air waveguide for different parameter sets. Solid line is for normalized $\beta$ of the guided modes. Dashed line is for normalized $\alpha$ of the complex guided modes. Blue color is for the forward ordinary modes; red color is for the backward ordinary modes; green color is for the complex guided modes. The green circles represent degeneracy points of forward-wave and backward-wave modes. The insets show the magnetic field patterns ($H_y$) for different modes. The $H_y$ field patterns in (b), (c) is similar with that in (a), and those in (e), (f) is similar with that in (d).

It is clear from the dispersion relations that TM mode solutions are closely related with the values of $\varepsilon_x$, $\varepsilon_z$ and $\mu_y$. For oscillating modes with real propagation constant, $p$ and $\kappa$ should be real, so that the parameters should be restricted by $0 < n_{eff} < 1$ and $n_{eff}^2 \varepsilon_z/\varepsilon_x > \varepsilon_z \mu_y/\varepsilon_0 \mu_0$. Table 1 summarizes the six possible sets of material parameters with which TM oscillating modes could exist. The complete dispersion diagrams belong to different sets, including ordinary and complex guided modes, are illustrated in Fig. 2. It shows that the waveguide could present various wave propagating properties under different cases. The notation "TM" in Fig. 2 identifies the polarization, followed by a subscript to track the number of field nodes in the core region and a superscript "$f$" or "$b$" to designate whether the mode is forward or backward, i.e., $TM_{m+1}^f$ for the $(m+1)^{th}$-order forward mode. The insets in Fig. 2 show the corresponding field patterns for different modes.

When $\varepsilon_x < 0$ (Fig. 2(c), 2(e), 2(f)), one may notice that some of the modes have forward wave branch, as well as backward wave branch with antiparallel phase ($v_p$) and group ($v_g$) velocities. The two branches merge at a mode-degeneracy point, corresponding to a 'critical' reduced slab thickness with zero group velocity. Due to the difficult to achieve magnetic response at optical range, we restrict on non-magnetic AMM with permeability being 1. Therefore, we focus on Set E. Corresponding to the parameters used in Fig. 2(e), Fig. 3(a) and 3(b) denote the variation of the group velocity and the normalized power flow $P = P_{tot}/(|P_1|+|P_2|+|P_3|)$ with reduced slab thickness. It is clear that the total power $P_{tot}$ vanishes and the group velocity reduces to zero at the corresponding degeneracy points. For complex guided modes, the total power flow and the group velocity are both zero, as we predicted previously.

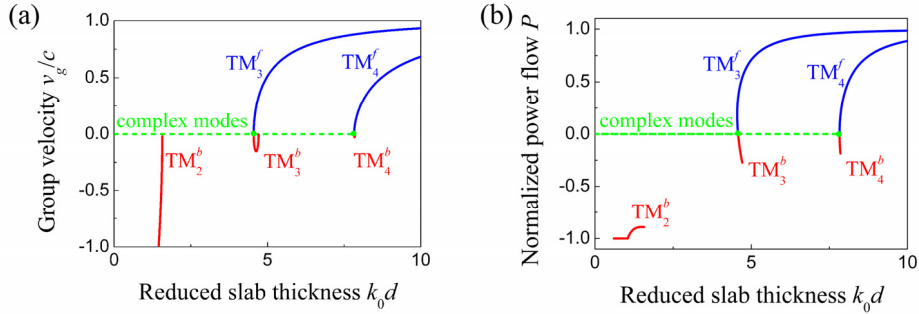

Fig. 3 Variation of group velocity (a), and normalized power flow (b) with reduced slab thickness for $\varepsilon_x = -0.1$, $\varepsilon_z = -1$ and $\mu_y = 1$. Blue color is for the forward ordinary modes; red color is for the backward ordinary modes; green color is for the complex guided modes. The green circles denote the degeneracy points.

## 3. Slow light waveguide analysis and simulation

The unique properties of the propagating modes can be used to slow down and even trap light. In Fig. 3(b), the group velocity decreases while the slab thickness reduces to the critical thickness for $TM_3$ or $TM_4$ mode at certain excitation frequency. Therefore, we design a linearly tapered waveguide to stop the wave, as illustrated in Fig. 4(a), with the same material parameters used in Fig. 3. Here we choose $TM_3^f$ mode as the operating mode because single-mode propagating can be excited while the air core thickness is correctly designed. FDTD method is employed to simulate the light propagating in the structure. Figure 4(b) demonstrates snapshots of the propagation of a monochromatic Gaussian modulated pulse with p-polarized magnetic-field component (with wavelength of 600nm) in the structure at different moments. The light pulse enters the air waveguide from the wide end (the width of air core is 1.4μm). The width of the other end is 200nm, which is less than the 'critical' thickness (435nm). The AMM in our simulation is modeled by a lossless, dispersive material obeying Drude model with $\varepsilon_x = -0.1$, $\varepsilon_z = -1$ and $\mu_y = 1$ at 600nm wavelength. While the guided oscillating fields propagate along the tapered waveguide, the group velocity progressively decreases. At a certain moment, the front part of the light pulse always travels slower than the rear part; therefore the pulse exhibits a spatial compression due to the reduction in group velocity while propagating along the structure, as demonstrated in Fig. 4(b). Eventually, the group velocity becomes zero at the 'critical', pre-determined, waveguide thickness (around 80T in Fig. 4(b)). At that position, the pulse is extremely compressed into a small region and stays there. Meanwhile, the amplitudes of the field components increase to about 4 times of the input values. The stopped light pulse should stay there for a long time. However, due to the variation of the core thickness *d* with distance *z* is not adiabatic, we can not expect to observe the splitting of the pulse in its frequency constituents and "stopping" of each frequency at a different location [21]. The amplitude of the trapped light will progressively decrease since it excites a backward-propagating forward-mode after staying for a while and the energy gradually spread out to the input end and fade out. If we try to totally stop a light, a very long adiabatically tapering structure should be used, which means that it is not easy to realize a perfect trap of the light. It should be noticed that this is totally different from the reflection by a cut-off ordinary waveguide since the light spreads out not in a pulse form as what is reflected in a cut-off case.

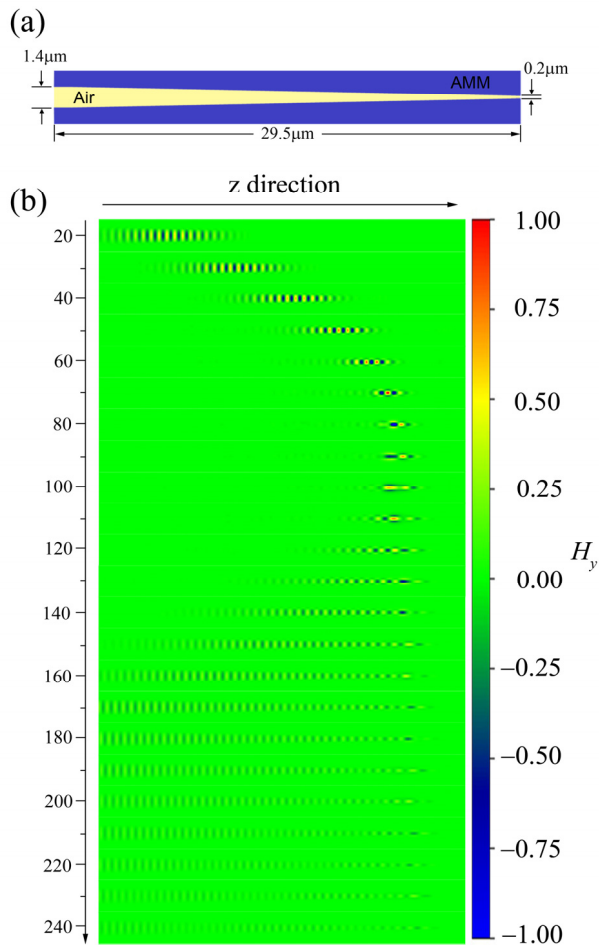

Fig. 4 (a) Geometry of the linearly tapered air waveguide with anisotropic metamaterial cladding. The material parameters of the AMM are chosen as $\varepsilon_x = -0.1$, $\varepsilon_z = -1$ and $\mu_y = 1$. (b) Snapshots of the propagation of a monochromatic (wavelength is 600nm) Gaussian modulated pulse (p-polarized magnetic-field component). The duration of the pulse is $5e^{-14}$ s. The time interval between every two snapshots is 10T (T is the period of the input light)(Media 1).

From previous analysis, we believe that for different excitation frequencies the guided lights will stop correspondingly at different guide thicknesses in the trapper. FDTD simulations are performed to verify this analytical result (Fig. 5). It shows that the full-wave simulation agrees with the analytical result very well. The propagating light having higher frequency stops at the position with smaller waveguide thickness. If the incident light is a white light, it can be altogether trapped within a fixed area, spanning a continuous range of waveguide thicknesses.

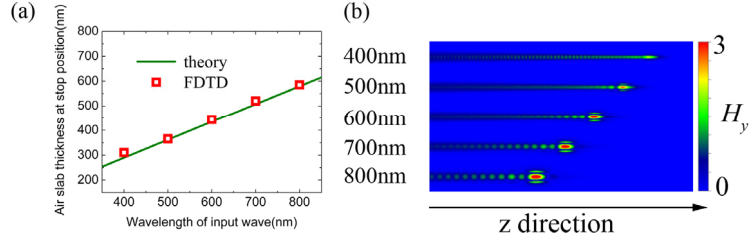

Fig. 5 (a) The relationship between the air slab thicknesses of the stop position and the excitation frequencies (material dispersion is not considered here). (b) Spatial field distributions in the tapered waveguide for light launched from the wider port, with the corresponding wavelengths marked on the left side.

Considering practical realization, we propose a layered medium (shown in Fig. 6), consisting of alternating silicon dioxide layer ($\varepsilon_1 = 3.9$) and silver layer ($\varepsilon_2 < 0$ at optical frequency), which can be effectively mimicked as an anisotropic metamaterial [15,23]. The effective dielectric permittivity tensor of such structure (with $N$ as the volume fraction of silver) is given by [24]

$$\varepsilon_x = \varepsilon_y = (1-N)\varepsilon_1 + N\varepsilon_2$$
$$\varepsilon_z = \frac{\varepsilon_1 \varepsilon_2}{N\varepsilon_1 + (1-N)\varepsilon_2} \quad . \tag{3}$$

At optical frequency silver behaves as a plasmonic material with $\varepsilon_2 < 0$. According to the parameter data in Ref. 25, we have for example, $\varepsilon_x = -0.125$, $\varepsilon_z = -3.31$ with the silver volume fraction $N = 70\%$ at 351nm. Thus by fabricating a light trapper with such multilayered composite material, we could probably stop the 351nm light at the position when the waveguide thickness reaches 292nm.

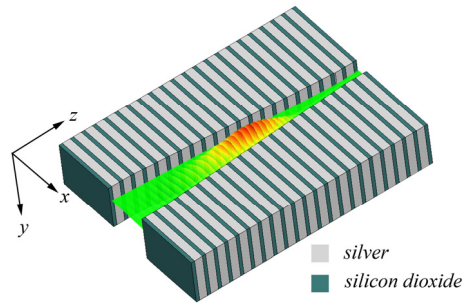

Fig. 6 A proposed tapered waveguide fabricated with aniostropic metamaterial, which is composed of a stack of alternating layers of silver and silicon dioxide, with layer thickness much less than the wavelength.

## 4. Conclusion

In summary, we have investigated the guided modes in a planar air waveguide with AMM cladding. It is shown that both forward and backward waves can propagate under certain conditions, differing considerably from those of a conventional dielectric waveguide. The propagating electromagnetic waves can be stopped at the critical point where the forward and backward modes degenerate together, as a result of the power flows in the two media totally cancelling out. A linearly tapered air waveguide is introduced to bring the propagating wave to a complete standstill. We believe that such air waveguide have two distinguished advantages, i.e., less influence of material losses on the guiding light, and easier fabrication with layered metal/dielectric structures. Therefore, the proposed air waveguide is more suitable for most conceived applications.

**Acknowledgments**

This work is supported by the National Basic Research Program of China (No. 2004CB719800) and the National Nature Science Foundations (No. 60671002 and No. 60801001).